\documentclass[aps,twocolumn,secnumarabic,superscriptaddress,balancelastpage,amsmath,amssymb,nofootinbib,titlepage]{revtex4}
\usepackage{amssymb}
\usepackage[english]{babel}
\usepackage[utf8]{inputenc}
\usepackage{amsmath}\usepackage{bbold}
\usepackage{graphicx}
\usepackage[colorinlistoftodos]{todonotes}
\usepackage{bibentry}
\usepackage{url}
\usepackage{rotating}
\usepackage{float}
\usepackage{color}
\usepackage[sort&compress]{natbib}
\usepackage{hyperref}
\usepackage{empheq}
\usepackage{textcomp}
\usepackage[caption=false]{subfig}
\usepackage{graphicx}

\newcommand{\bone}{\mathbb{1}}

\newcommand{\curP}{{\cal P}}
\newcommand{\nf}{n_{\rm F}}
\newcommand{\kb}{k_{\rm B}}
\newcommand{\Ebar}{\bar{E}}

\newcommand{\psih}{\hat{\psi}}
\newcommand{\phdag}{{\phantom{\dagger}}}

\newcommand{\Psih}{\hat{\Psi}}
\newcommand{\chih}{\hat{\chi}}

\newcommand{\sgn}{{\rm sgn}}

\newcommand{\be}{\begin{equation}}
\newcommand{\ee}{\end{equation}}
\newcommand{\bea}{\begin{eqnarray}}
\newcommand{\eea}{\end{eqnarray}}
\newcommand{\bse}{\begin{subequations}}
\newcommand{\ese}{\end{subequations}}

\begin{document}
\title{Analog Unruh effect of inhomogeneous one-dimensional Dirac fermions}

\author{Khristian B. Tallent}
\email{tallent9997@gmail.com}
  \author{Daniel E. Sheehy}
\email{sheehy@lsu.edu}
  \affiliation{Department of Physics and Astronomy, Louisiana State University, the Baton Rouge, LA 70803 USA}

\date{Aug. 1, 2024}

\begin{abstract}
We study one-dimensional Dirac fermions in the presence of a spatially-varying Dirac velocity $v(x)$,
that can form an approximate lab-based Rindler Hamiltonian describing an observer
accelerating in Minkowski spacetime.  A sudden switch from a spatially homogeneous velocity 
($v(x)$ constant) to a spatially-verying velocity ($v(x)$ inhomogeneous) leads to 
the phenomenon of particle creation, i.e., an analog Unruh effect.  
We study the dependence of the analog Unruh effect on the precise form of the velocity profile, 
finding that while the ideal Unruh effect occurs for $v(x)  \propto |x|$, a modified Unruh effect
still occurs for more realistic velocity profiles that are linear for $|x|$ smaller than a length scale 
$\lambda$ and constant for $|x|\gg \lambda$ (such as $v(x)\propto \tanh \big(|x|/\lambda \big)$).  We show that the 
associated particle creation is localized to $|x|\ll \lambda$.  
\end{abstract}

\maketitle 

\section{Introduction}
\label{sec:intro}
The field of analog gravity began with the proposal, by Unruh, that sound waves 
traveling in moving fluids 
 could mimic the phenomenon of black hole evaporation~\cite{Unruh:1980cg}.  
The idea of analog gravity (reviewed in Refs.~\cite{Barcelo:2005fc,Nation:2011dka,Jacquet2020}) is that by studying 
lab-based systems obeying similar equations to those of astrophysical
or relativistic 
systems of interest, we can better understand the latter in a controlled experimental setting.  This field can also stimulate novel understanding by bringing ideas from curved spacetime quantum field theory into 
condensed-matter  and cold-atom settings.
Some of the analog gravity platforms that have been studied include
proposals to simulate black holes and Hawking Radiation~\cite{Philbin:2007ji,Belgiorno:2010wn,Weinfurtner:2010nu,Steinhauer:2015saa}, the properties of the expanding universe
such as inflation~\cite{Fischer:2004bf,Prain:2010zq, Eckel:2017uqx,Wittemer2019,Banik:2021xjn,
Llorente:2019rbs,
Bhardwaj:2020ndh,Bhardwaj2024}
 and curvature~\cite{Viermann:2022wgw,Tolosa-Simeon:2022umw,SK2022},
 and other phenomena such as 
 the dynamical Casimir effect~\cite{Jaskula2012},
Sakharov oscillations~\cite{Hung:2012nc}, and 
rotational superradiance~\cite{Delhom}.

Here, we analyze a system of Dirac fermions in one spatial 
dimension, that can exhibit  an analog Unruh effect~\cite{Fulling:1972md,Davies:1974th,Unruh:1976db}.  Indeed, there
have been numerous studies of analog systems for the Unruh effect,
including in 
in cold atomic gases 
\cite{Retzker:2008,Boada:2010sh,Rodriguez-Laguna:2016kri,Kosior:2018vgx,Hu:2018psq,Gooding:2020scc},
graphene~\cite{Iorio:2011yz,Iorio:2013ifa,Cvetic:2012vg,Bhardwaj2023}, 
Weyl semimetals~\cite{Volovik:2016kid}, and 
quantum hall systems~\cite{Hegde:2018xub,Subramanyan:2020fmx}.
In the astrophysical context, the Unruh effect describes how
an accelerating observer will measure
a thermal distribution of particles even if a stationary observer
measures the vacuum state~\cite{Fulling:1972md,Davies:1974th,Unruh:1976db}.  The corresponding temperature of
such particles, $T_{\text{U}}=\frac{\hbar a}{2\pi k_{\text{B}} c}$,
is proportional to the acceleration $a$ of the observer, with
$k_{\text{B}}$ the Boltzmann constant, $\hbar$ Planck's constant and $c$ the speed of light.  

The origin of the Unruh effect (reviewed in Ref.~\cite{Crispino:2007eb})
can be traced to the fact that the spacetime
 for accelerating observers, the Rindler spacetime~\cite{Rindler:1966zz},
effectively splits the universe into two mutually inaccessible regions. 
From a quantum field theory perspective, these two regions each have
their own mode operators describing particle excitations that are 
non-trivially related to mode operators of the static Minkowski vacuum.  The pure Minkowski vacuum
appears mixed to the Rindler observers, leading to thermal expectation values for 
observables.

\begin{figure}[h]
\centering
\includegraphics[width=\columnwidth]{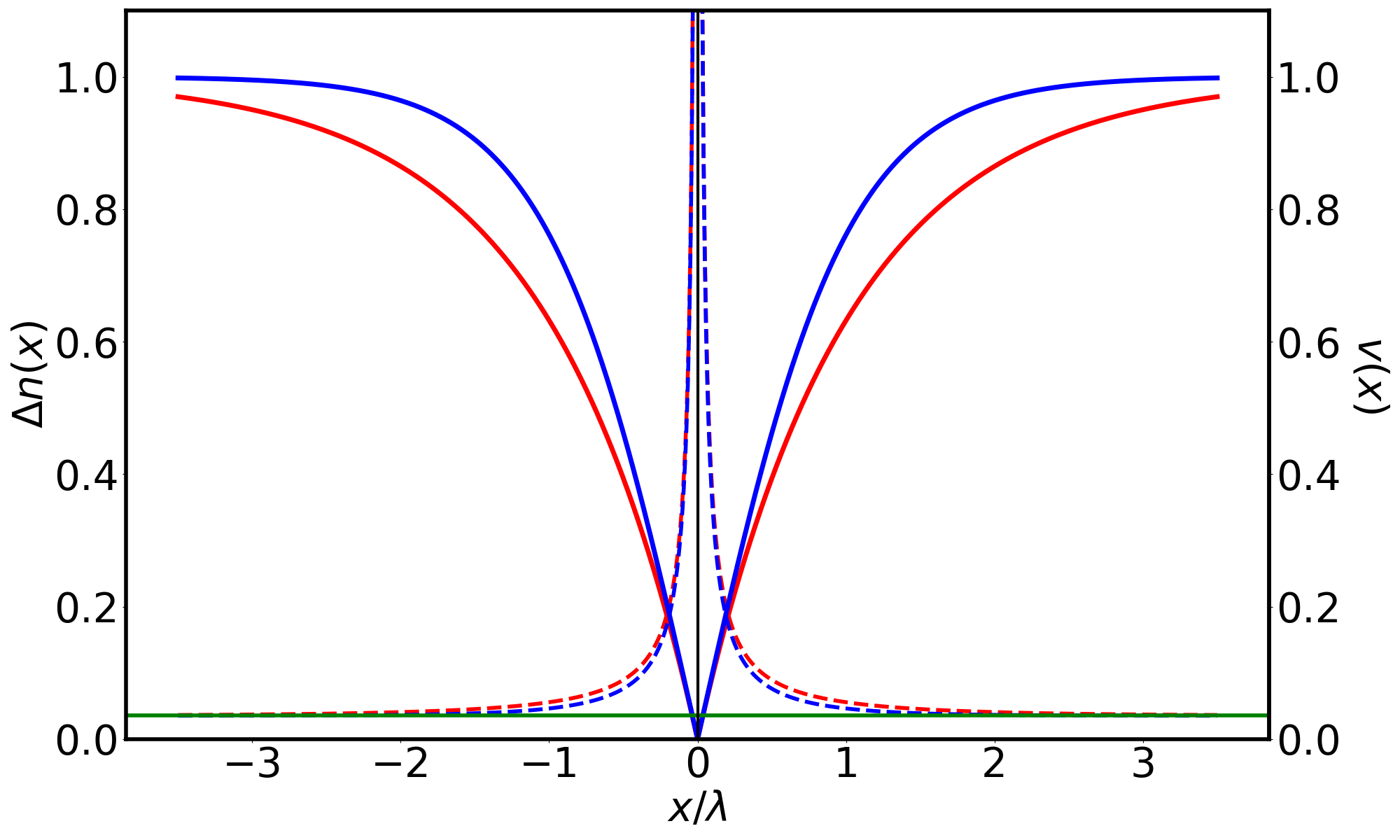}
\caption{The solid curves show the normalized inhomogeneous velocity profiles for a 1D Dirac system, 
$v(x) \propto \tanh |x|/\lambda$ (tanh case, upper blue curve), and
$v(x) \propto (1-{\rm e}^{-|x|/\lambda})$ (sigmoid case, lower red curve) and 
characterized by a length scale $\lambda$.
The dashed curves show the predicted local densities of positive-energy 
fermions induced after a sudden quench from the homogeneous
velocity case to the inhomogeneous profile, representing an analog
Unruh effect.
Here the upper red dashed curve
(lower blue dashed curve) corresponds to the sigmoid (tanh) cases, with the green curve being the asymptotic
value at large $|x|$.  
} 
\label{fig:LFD}
\end{figure}

To achieve an analog Unruh effect in a lab setting, it is not 
necessary to engineer an accelerating Rindler observer.  Rather, a simpler
strategy is exploit the fact that a coordinate change 
transforms the Rindler metric to that of a spacetime with a 
spatially-varying speed of light, of the form $v(x) \propto |x|$, which plays the role of
the Rindler observer in the analog setup (as we show below). 
However, this leads to additional difficulties, since it is 
challenging to realize a lab-based system described by a Dirac equation
with $v(x) \propto |x|$ for all $x$.  For example, it is well-known that
graphene-like systems with spatially-varying nearest-neighbor tunneling
(realizable with a spatially-varying imposed strain~\cite{deJuan:2012hxm}
or with an appropriately engineered light field~\cite{Jimenez-Galan:2020}) 
can yield a spatially varying velocity.  However, it is difficult to 
see how such approaches can lead to $v(x) \propto |x|$ for all $x$ (a
highly nonperturbative change in the local Dirac velocity).  

A natural next question, to be addressed here, is to what extent
an approximate 
analog Unruh effect can be realized in a Dirac-type fermion system with
a more realistic localized velocity profile, such as $v(x)  = v \tanh(|x|/\lambda)$.
Such a velocity profile, illustrated as the solid (upper) blue line in Fig.~\ref{fig:LFD},
exhibits $v(x)\propto |x|$ for $|x|\ll \lambda$, with
$\lambda$ a length scale characterizing the size of the distortion.  For $|x|\gg
\lambda$, we have $v(x) = v$, the background velocity in the absence of the applied strain.  Although easier to realize in an experiment, such a
velocity profile can only yield an approximate Unruh effect, since the
mapping to a uniformly accelerating observer no longer holds.  
Despite this,  we find the approximate Unruh effect to still be characterized by the same Unruh 
temperature as the strictly linear case $v(x)\propto |x|$. 
We find that the positive-energy
fermions associated with this approximate emergent Unruh effect are spatially 
localized near $x=0$, as seen in the dashed lines of Fig.~\ref{fig:LFD}.

This paper is organized as follows.   
In Sec.~\ref{sec:model} we describe our model Hamiltonian for
a 1D Dirac system with an inhomogeneous local Dirac velocity and review
the system ground state for the well-known case of a uniform velocity.
In Sec.~\ref{sec:linear} we review how a quench from an initial 
uniform-velocity case to the linear velocity case $v(x)\propto |x|$
yields the Unruh effect: an emergent Fermi distribution of particles despite the
physical system being at zero temperature.   In Sec.~\ref{sec:nonlinear},
we generalize these calculations to the case of the 
nonlinear velocity profiles shown in Fig.~\ref{fig:LFD}, finding
an approximate Unruh effect still emerges in this case.  
In Sec.~\ref{sec:extent}, we study the spatial distribution of
positive-energy fermions associated with this Unruh effect, showing
the spatial distributions given by the dashed cuves in Fig.~\ref{fig:LFD}.
In Sec.~\ref{sec:concl} we provide brief concluding remarks and
discuss future directions for study.

\section{Model Hamiltonian}
\label{sec:model}
Following the approach of Ref.~\cite{Bhardwaj2023}, which analyzed
the Unruh effect for 2D Dirac systems with a spatially-varying velocity,
here we analyze a 1D version with Hamiltonian 
\begin{gather}
\label{eq:flatHam}
 H = \sqrt{v(x)}\sigma_{x} p_x \sqrt{v(x)},
\end{gather}
where $p_{x}=-i\hbar\partial_{x}$ is the usual $x$-momentum operator
and $\sigma_{x}=\big(\begin{smallmatrix} 0 & 1\\ 1 & 0 \end{smallmatrix}\big)$ is the Pauli matrix.  
We note that, as written, the Hamiltonian in Eq.~(\ref{eq:flatHam}) is 
Hermitian, and that below  Planck's
constant $\hbar$ will generally be set to unity.
Before proceeding, we remark that
inhomogeneous Dirac systems like Eq.~(\ref{eq:flatHam}) have been studied
in other contexts, for example Refs.~\cite{Ghorashi:2020,Davis:2022} that studied 
analog gravitational lensing   and Ref.~\cite{Morice2022} that studied quantum dynamics
in lattice models that realize Hamiltonians like Eq.~(\ref{eq:flatHam}).

Our first task is to recall the case of 
a spatially-uniform velocity, $v(x)=v$.  In this case, 
Eq.~(\ref{eq:flatHam}) is a model for 1D Dirac Fermions
with eigenfunctions $\hat{\psi}_{p\alpha}(x) = \hat{\chi}_\alpha {\rm e}^{ipx}$
and energies $E_{\alpha}(p) = \alpha vp$ where $\alpha = \pm$.
Here and below hatted quantities denote two-component spinors with 
components labeled by $i,j$.  
The spinors $\hat{\chi}_{\alpha}$
are just the eigenspinors of $\sigma_x$, explicitly given by 
$\hat{\chi}_{+}=\frac{1}{\sqrt{2}}\big(\begin{smallmatrix} 1\\ 1 \end{smallmatrix}\big)$ and $\hat{\chi}_{-}=\frac{1}{\sqrt{2}}\big(\begin{smallmatrix} 1\\ -1 \end{smallmatrix}\big)$.

Turning to the many-body case with spinless fermions, 
we define two-component real-space field 
operators $\Psih(x)$ obeying the anticommutator relation 
\be
\label{Eq:anticommutator}
\{\Psi_i(x),\Psi_j^\dagger(x')\} = \delta_{ij}\delta(x-x').
\ee
%
%
To describe the system ground state and excitations, we define mode operators $c_{p\alpha}$ ($c_{p\alpha}^\dagger$) that 
annihilate (create) particles in the single-particle state $\hat{\psi}_{p\alpha}$.
These operators obey the anticommutator relation 
\be
\{c_{p\alpha}^\phdag,c_{p'\beta}^\dagger\} = 2\pi\delta_{\alpha\beta} \delta(p-p').
\ee
The field and mode operators are connected by the relations 
\bea
\label{Eq:psihc}
 \Psih(x) &=& \sum_{\alpha = \pm} \int_{-\infty}^\infty \frac{dp}{2\pi}\, \psih_{p\alpha}(x)c_{p\alpha},
 \\
 c_{p\alpha} &=&
\int_{-\infty}^\infty dx \,\psih^{\dagger}_{p\alpha}(x)\Psih(x),
\eea
Having defined the relevant quantum field and mode operators, next
we define the system ground state (the analog Minkowski vacuum), in which all negative (positive) energy states
have unit (vanishing) occupation.  This is captured by the expectation value (EV):
\be
\label{Eq:MV}
\langle c_{p\alpha}^\dagger 
c_{p'\beta}^\phdag \rangle = 2\pi\delta_{\alpha\beta}\Theta(-\alpha p)
\delta(p-p').
\ee
The Unruh effect emerges when the Minkowski state is \lq\lq observed" by the
Rindler system, which consist of Dirac fermions in the presence of an engineered
velocity profile $v(x)$.  

The proposed experimental procedure to realize this is as 
follows~\cite{Bhardwaj2023}:  A system of Dirac fermions 
obeying Eq.~(\ref{eq:flatHam})
is prepared in the Minkowski
vacuum Eq.~(\ref{Eq:MV}).  Subsequently, a rapid quench to the engineered
profile $v(x)$ is induced, for example by imposing a  strain~\cite{deJuan:2012hxm}
or an external light field~\cite{Jimenez-Galan:2020},
which can modify the hopping matrix elements of an underlying tight-binding model and hence the local velocity.
 Within the sudden approximation of quantum 
mechanics, observables in the final system are computed by
taking the EV of the corresponding operators with respect
to the initial (pre-quench) state, i.e., the Minkowksi vacuum.  
The emergent Unruh effect reflects the fact that this final EV
describes spontaneous particle creation, as we shall see.

Having defined the initial Minkowski state in this section, our next task is to analyze possible final states, described by a Dirac equation with a spatially-inhomogeneous velocity profile.

\section{Linear Velocity Profile: Ideal Unruh effect}
\label{sec:linear}
As we have discussed, our aim is to investigate the emergence of the Unruh effect
in a simple 1D Dirac model with a spatially-varying velocity $v(x)$.
In particular, we are interested in how the associated particle
creation depends on the details of $v(x)$.  In this section, we study the
case of a linear velocity profile, $v(x) = v|x|/\lambda$.  We emphasize
that such a linear profile corresponds, in the analogy to the
conventional Unruh effect, to a uniformly accelerating Rindler observer.
Thus, we expect (and indeed find) an exact Unruh effect in this case.  
This case will serve as a basis for what to expect for the subsequent 
velocity profiles, discussed below.  

To demonstrate the Unruh effect in this system, we first compute the eigenfunctions of $H$, Eq.~(\ref{eq:flatHam}), for the present velocity
profile. The corresponding Hamiltonian is:
\begin{gather}
\label{Eq:hamstrained}
    H = \frac{v}{\lambda}\sqrt{|x|}\sigma_x (-i\hbar \partial_{x})\sqrt{|x|},
\end{gather}
with $\lambda$ a length scale associated with the spatial variation
of the velocity profile.  At this point we set $\lambda = 1$,
$v=1$ and $\hbar =1$ for simplicity, restoring them at the end of the section.

A key property of this particular Hamiltonian is the disjunction at $x=0$, thus rendering a spatially inhomogenous differential equation. This means we can treat the solutions as referring to their own half-space, separately
considering $x>0$ and $x<0$.   Focusing on $x>0$ (indicated with a subscript $>$), we find the eigenfunctions (with 
eigenvalue $E$)
\be
\psih_{>E\alpha}(x) = \frac{1}{\sqrt{2\pi x}}{\rm e}^{\alpha
  iE\ln x} \chih_{\alpha},
\ee
which satisfy orthonormality and completeness relations:
\bea
&&\int_0^\infty dx \, \psih_{>E\alpha}^\dagger(x)\psih_{>E'\beta}(x)  =
\delta_{\alpha\beta}\delta(E-E'),
\\
&& \sum_{\alpha = \pm} \int_{-\infty}^\infty dE\, \psih_{>E\alpha}(x)
\psih_{>E\alpha}^\dagger(x') = \delta(x-x')\bone,
\eea
where $\bone$ is the unit matrix in the two component
space of eigenfunctions of $H$.

As in the uniform $v$ case,
to study a many-body system of fermions described by the Hamiltonian 
Eq.~\ref{Eq:hamstrained} we introduce \lq\lq final\rq\rq\ mode operators $d_{>E\alpha}^\phdag$ ($d_{>E\alpha}^\dagger$) that annihilate 
(create) fermions in the state with energy $E$ and band index $\alpha$ in the $x>0$ region.  
Then, operators corresponding to observables in the final system can be built from these mode operators.  To apply the sudden approximation as
described above, in which any EVs should be computed with respect to the initial ground state, we must find a relation
between the final ($d_{>E\alpha}^\phdag$) and initial ($c_{p\alpha}^\phdag$) mode operators.  To achieve this, we first express
the real-space field operator (for $x>0$) in terms of the final mode operators:
\be
\label{eq:fieldOpform}
\Psih(x)  =  \sum_{\alpha = \pm} \int_{-\infty}^\infty dE 
\psih_{>E\alpha }(x) d_{>E\alpha}.
\ee
With this definition, the anti-commutation relation 
$\{ d_{>E_1\alpha}, d_{>E_2\beta}^\dagger\} = \delta(E_1-E_2) \delta_{\alpha\beta}$ is consistent with the real-space anti-commutator relation for the field
operators in Eq.~(\ref{Eq:anticommutator}).  The inverse relation 
to Eq.~(\ref{eq:fieldOpform}) is 
\be
\label{Eq:deePSI}
d_{>E\alpha} = \int_{0}^\infty dx \, \psih_{>E\alpha}^\dagger \Psih(x) .
\ee
Now, to achieve the desired expression of the $d_{>E\alpha}$ in terms of the $c_{p\alpha}$ we just need to insert Eq.~(\ref{Eq:psihc}) into the right side of 
Eq.~(\ref{Eq:deePSI}) and evalute the resulting $x$ integral.  We get:
\bea
    d^{\phdag}_{>E\alpha}\!\!&=&\!\!\sum_{\beta=\pm}\int_{-\infty}^{\infty}\frac{dp}{2\pi}\int^{\infty}_{0}dx\,\psi^{\dagger}_{>E\alpha}(x)\psi^{\phdag}_{p\beta}(x)c^{\phdag}_{p\beta},\\
        &=&\int_{-\infty}^{\infty}\frac{dp}{2\pi}X^{\phdag}_{\alpha}(p, E)c^{\phdag}_{p\alpha},
        \label{Eq:deecee}
\eea
where the function $X_{\alpha}^{\phdag}$ comes from the inner product of the strained and unstrained eigenfunctions
and is given by:
\bea
\label{Eq:exExpression}
&&X^{\phdag}_{\alpha}(p, E)= \int_{0}^\infty dx \frac{1}{\sqrt{2\pi x}}
    {\rm e}^{-i\alpha E\ln x}{\rm e}^{ipx},
    \\
 &&=    \frac{1}{\sqrt{2\pi |p|}} {\rm e}^{i\alpha E\ln |p|}
      {\rm e}^{i\frac{\pi}{4}{\rm sgn}(p)} {\rm e}^{{\rm sgn}(p)\alpha E\pi/2}
      \Gamma\Big[\frac{1}{2}-i\alpha  E\Big],
        \nonumber 
\eea
with $\Gamma(z)$ the gamma function.  
Then, given an initial state  in terms of the $c_{p\alpha}$ (e.g., a vacuum or thermal state), Eq.~(\ref{Eq:deecee}) allows
us to evaluate EVs of the final system in terms of the initial EV.  
For example, consider the $+$ band average 
\bea
&&\hspace{-0.5cm}
\langle d_{>E_1+}^{\dagger}  d_{>E_2+}^{\phdag} \rangle
\\
&& = \int_{-\infty}^{\infty}\frac{dp}{2\pi}X^{*}_{+}(p, E_1)
 \int_{-\infty}^{\infty}\frac{dp'}{2\pi}X^{\phdag}_{+}(p', E_2)
 \nonumber
\langle  c^{\dagger}_{p+},
 c^{\phdag}_{p'+}\rangle .
\eea
If we assume the Minkowski vacuum initial state, then
the EV on the right is given by Eq.~(\ref{Eq:MV}) above.   
Since we're studying the $\alpha = +$ case, with 
the $p<0$ states occupied and $p>0$ states empty, we get:  
\bea
\langle d_{>E_1+}^{\dagger}  d_{>E_2+}^{\phdag} \rangle &=& 
 \int_{-\infty}^{0} \frac{dp}{2\pi} X_{+}^*(p,E_1) X_+(p,E_2)
 \label{Eq:integralx1}
 \\
 &=&
  \delta(E_1-E_2) \frac{1}{{\rm e}^{2\pi E_1}+1},
\eea
i.e., a Fermi distribution at \lq\lq temperature\rq\rq\ $1/(2\pi)$.  
Following the same steps for the $-$ band we get:
\bea
 \langle d^\dagger_{>E_1-}  d_{>E_2-}\rangle
 &=&  \int_{0}^{\infty} \frac{dp}{2\pi} X_{-}^*(p,E_1) X_-(p,E_2),
 \label{Eq:integralx2}
 \\
 &=& 
 \delta(E_1-E_2) \frac{1}{{\rm e}^{2\pi E_1}+1},
\eea
the same final result again reflecting a Fermi distribution.
Taking these results together, and reintroducing
dimensionful parameters, we obtain (defining
the Fermi distribution 
$\nf(E,T) = \frac{1}{{\rm e}^{E/(\kb T)} +1}$):
\begin{gather}
\label{Eq:unruhAverage}
       \langle d_{>E_1\alpha}^{\dagger}d^{\phdag}_{>E_2\beta} \rangle=  \delta_{\alpha\beta}\nf(E_1, T_{\rm U})\delta(E_1-E_2),
\end{gather}
thereby recovering the well-known Unruh effect (also known
as the Fulling-Davies-Unruh effect~\cite{Fulling:1972md,Davies:1974th,Unruh:1976db}), with
the final  
Unruh temperature (restoring $v$, $\lambda$ and $\hbar$ that were previously set to unity):
\be
\label{eq:unruhTemp}
 T_{\rm U}=\frac{v\hbar}{2\pi\lambda \kb}.
\ee
Within this analog Unruh effect, the role of 
\lq\lq acceleration\rq\rq\ is determined
by the ratio $a = c^2/\lambda$, as can be seen by comparing to the 
conventional Unruh temperature 
$T_{\text{U}}=\frac{\hbar a}{2\pi k_{\text{B}} c}$ and identifying $v=c$.

Here we make a  technical remark that the final integrals in Eqs.~(\ref{Eq:integralx1}) and (\ref{Eq:integralx2}) can be most easily evaluated by
using the integral representation of the function $X_\alpha(p,E)$ 
in the first line of Eq.~(\ref{Eq:exExpression})
and evaluating the $p$ integrals first.  In the next section, in which we study
alternate velocity profiles $v(x)$, we will follow this strategy.

The results of this section confirm that an analog Unruh effect can emerge in a 1D Dirac model under the condition of a 
sudden quench of the velocity profile from a uniform homogeneous velocity $v(x) = v$ to a highly inhomogeneous velocity
profile $v(x)  =v|x|/\lambda$.  Our next task is to study
alternate velocity profiles in which the modification of
the velocity is spatially localized.

\section{Nonlinear velocity profile}
\label{sec:nonlinear}
As we have seen, the linear case gave the 
expected perfect Unruh effect, with the occupation of
states of the final system being given by a 
Fermi distribution.  However, it may be difficult
in a real experiment 
to realize the required perfect linear velocity profile 
for all $x$.  To investigate this, in this section
we consider nonlinear velocity profiles after the quench,
starting with a \lq\lq sigmoid\rq\rq\ velocity profile
followed by a \lq\lq tanh\rq\rq\ velocity profile, which 
will yield qualitatively similar results.  

\subsection{Sigmoid velocity profile}
The sigmoid velocity profile we study takes the form:
\be
\label{eq:sigmoidvee}
v(x) = v\big(1-{\rm e}^{-|x|/\lambda}\big).
\ee
going as $|x|$ for $x\to 0$ but constant at large
$|x|\gg \lambda$, with $\lambda$ a length scale 
characterizing the size of the deformation.  Once again,
we take $v,\lambda \to 1$
for simplicity, only reintroducing them in final results.  Since $v(x)$ is only significantly modified near
$x =0$, achieving this case in an experiment is expected to be easier than in the $v(x) \propto |x|$ case.

As in the preceding section, we need the eigenfunctions of the single-particle Hamiltonian corresponding
to $v(x)$ in Eq.~(\ref{eq:sigmoidvee}):
\begin{gather}
    H = \sqrt{1-e^{-|x|}}{(-i\partial_{x})}\sqrt{1-e^{-|x|}}.
\end{gather}
It is sufficient to focus on $x>0$.  We find the eigenfunctions 
\be
 \psih_{>E\alpha}(x) = \frac{1}{\sqrt{2\pi}}
     {\rm e}^{\frac{1}{2}x} \big( {\rm e}^{x} -1\big)^{i\alpha
       E -\frac{1}{2}}\chih_\alpha,
     \label{eq:psiSigmoid}
\ee
that are orthonormal and complete in the region $x>0$.  As previously, to investigate the Unruh effect
we express the real-space field operators in terms of mode operators (again called 
$d_{>E\alpha}^\phdag$ and $d_{>E\alpha}^\dagger$).  These operators can again be connected to the initial mode operators via an 
expression of the form of Eq.~(\ref{Eq:deecee}) but with a different function $X_\alpha(p,E)$.  Although we can find an explicit 
expression for the corresponding function, it is easier to work with an integral representation 
when evaluating the number operator in the present case.

We focus on the $\alpha = +$ band with similar results
holding for the $-$ case.  Following similar steps to the 
previous section (assuming the initial system EV given by Eq.~(\ref{Eq:MV})), 
we find:
\bea
&&\hspace{-0.5cm}\langle d_{>E_1+}^\dagger d_{>E_2+}\rangle
=  \int_{-\infty}^{0} \frac{dp}{(2\pi)^2} 
\int_0^\infty dx_1 {\rm e}^{\frac{1}{2}x_1} \big( {\rm e}^{x_1} -1\big)^{i
  E_1 -\frac{1}{2}} \nonumber 
\\
&&\qquad \times 
\int_0^\infty dx_2 {\rm e}^{\frac{1}{2}x_2} \big( {\rm e}^{x_2} -1\big)^{-i
  E_2 -\frac{1}{2}} {\rm e}^{-ip(x_1-x_2)},
  \label{eq:sigmoidDaverage}
\eea
for the expectaton value of the final system mode operators.    

As mentioned above, we find it advantageous to first evaluate the $p$ integral, which we interpret in the sense of a 
distribution using: 
\bea
&&\int_{-\infty}^0 \frac{dp}{2\pi} {\rm e}^{-ip(x_1-x_2)}  = {\rm Lim.}_{\eta\to 0+} \int_{-\infty}^0 \frac{dp}{2\pi} {\rm e}^{-ip(x_1-x_2)} 
{\rm e}^{\eta p} \nonumber 
\\
&&\qquad \qquad =  {\rm Lim.}_{\eta\to 0+} \frac{1}{2\pi} \frac{1}{\eta -i(x_1-x_2)}\\
&& \qquad \qquad =
 \frac{i}{2\pi}\Big( \curP \frac{1}{x_1-x_2} -i\pi \delta(x_1-x_2)\Big),
 \label{Eq:spformula}
\eea
with $\curP$ indicating principal value and where we have used the Sokhotski-Plemelj formula.  Upon plugging this in
to Eq.~(\ref{eq:sigmoidDaverage}), the contribution from the delta function will pin $x_2=x_1$. Dropping the subscript,
the resulting $x$
integral gives an energy delta function: 
\be
\frac{1}{2\pi} \int_0^\infty dx\, {\rm e}^x \Big({\rm e}^x -1\Big)^{i(E_1-E_2)-1} =
  \delta(E_1-E_2),
  \label{eq:deltafunction}
\ee
which is easiest to check by changing variables to $y= \ln \big[{\rm e}^x-1\big]$,
leading to the well-known delta-function representation 
\be
\int_{-\infty}^\infty \frac{dy}{2\pi} {\rm e}^{iy(E_1-E_2)}= \delta(E_1-E_2).
\ee
The formula Eq.~(\ref{eq:deltafunction}) that simplified the delta-function contribution to Eq.~(\ref{eq:sigmoidDaverage})
 is in fact a re-statement of the orthonormality of the system eigenfunctions Eq.~(\ref{eq:psiSigmoid}).  Including the 
 contribution from the principal value term in Eq.~(\ref{Eq:spformula}), we have 
\bea
\label{eq.opavg+corr}
\langle d^{\dagger}_{>E_{1}+}  d_{>E_{2}+}^{\phdag}\rangle=\frac{1}{2\pi} \big( \frac{1}{2}\delta(E^{\phdag}_{1}-E^{\phdag}_{2})+\mathfrak{R}(E^{\phdag}_{1},E^{\phdag}_{2}) \big),
\eea
where the second term in parentheses is: 
\bea
&&\mathfrak{R}(E_{1}, E_{2}) 
 = 
 \frac{i}{4\pi^2}
\int_0^\infty dx_1 \,{\rm e}^{\frac{1}{2}x_1} \big( {\rm e}^{x_1} -1\big)^{i
  E_1 -\frac{1}{2}}  \nonumber 
\\
&&\times 
\int_0^\infty dx_2 \,{\rm e}^{\frac{1}{2}x_2} \big( {\rm e}^{x_2} -1\big)^{-i
  E_2 -\frac{1}{2}}  \curP\frac{1}{x_1-x_2} .
\label{eq:cor.term}
\eea

\begin{figure}[h]
\centering
\includegraphics[width=\columnwidth]{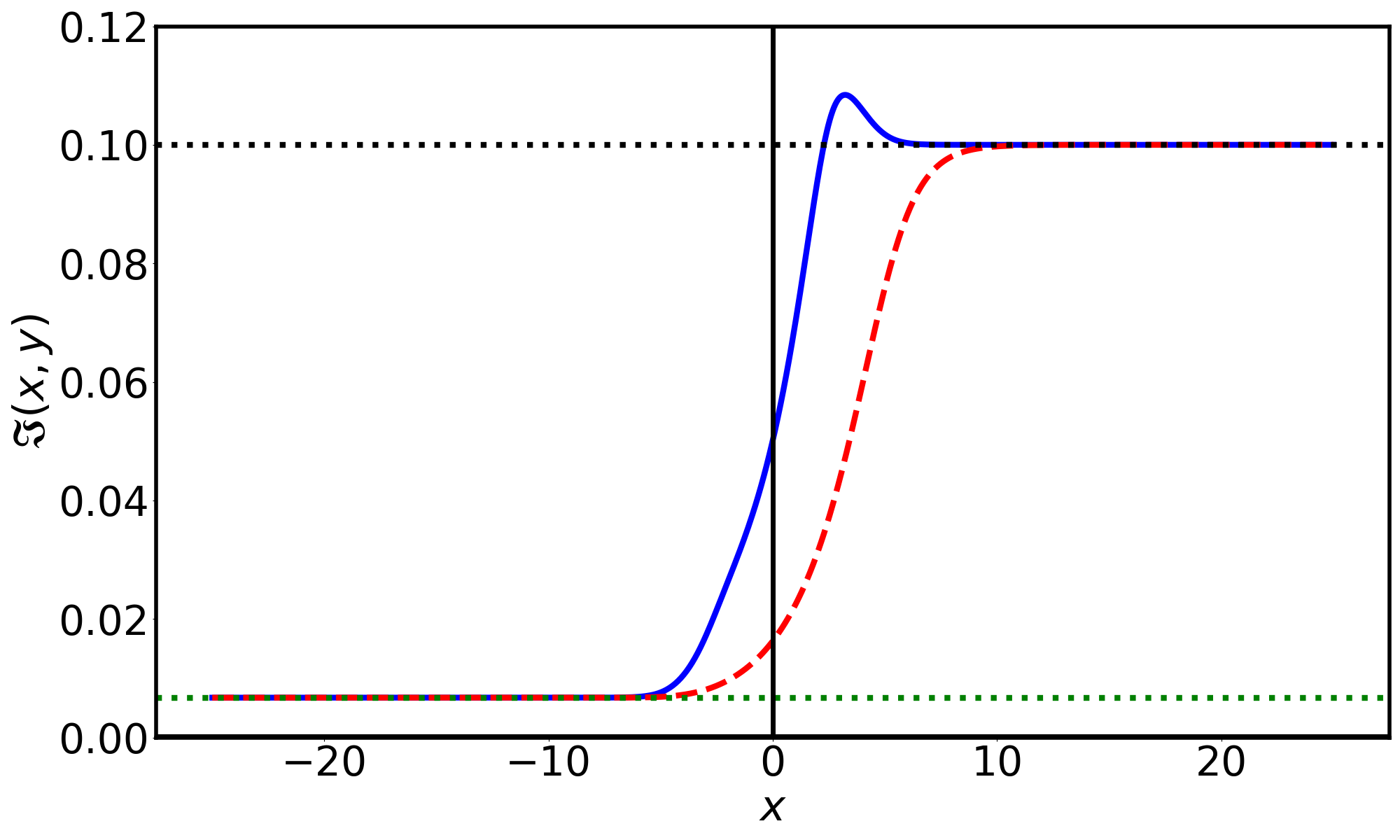}
\caption{Plots of the integrand $\mathfrak{I}(x,y)$ in the sigmoid case 
Eq.~(\ref{eq:isig})  (red dashed) and in the hyperbolic tangent case
Eq.~(\ref{eq:Itanh}) (solid blue) for the case $y=10$, showing
that this function rapidly reaches the asymptotic behavior given by 
Eq.~(\ref{eq:FutureReference}).  
To illustrate these limits, the top black dotted line is at $1/y=0.1$
and the bottom green dotted line is at $1/(2\sinh(y/2))\simeq 0.0067$.}
\label{IPlots}
\end{figure}
To study the contribution due to $\mathfrak{R}(E_{1}, E_{2})$, we make a variable change:
\begin{gather}
x^{\phdag}_{1}=\ln(e^{{x}+y/2}+1), x^{\phdag}_{2}=\ln(e^{{x}-y/2}+1).
\end{gather}
The resulting Eq.~(\ref{eq:cor.term}) becomes: 
\bea
\label{eq:approxR}
&&\mathfrak{R}(E_{1}, E_{2}) 
\\
&& = \frac{i}{4\pi^2} \int_{-\infty}^{\infty}dx\int_{-\infty}^{\infty}dy\,\mathfrak{I}(x, y)e^{i(E_{1}-E_{2})x}e^{\frac{i}{2}(E_{1}+E_{2})y}. \nonumber
\eea
where the integrand $\mathfrak{I}(x,y)$ is given by:
\be
\label{eq:isig}
\mathfrak{I}(x,y)\!=\! \sqrt{\frac{e^{x}}{2(\cosh x\!+\!\cosh(y/2))}}\curP\frac{1}{\ln\left( \frac{e^{x+y/2}+1}{e^{x-y/2}+1} \right)}.
\ee
Although the remaining integrals are too difficult to evaluate directly, we can make a simple approximation by noting that, as a 
function of $x$,  
$\mathfrak{I}(x,y)$ rapidly reaches an asymptotic $y$-dependent value for $x\to \pm \infty$.  Indeed, we find the limiting behavior
\be
\mathfrak{I}(x,y)  = 
\begin{cases}
\frac{1}{y},
& \text{for  $x\to +\infty$,}\cr
\frac{1}{2 \sinh(y/2)}
& \text{for $x\to -\infty$.}
\end{cases}
\label{eq:FutureReference}
\ee
This limiting behavior is easily verified by plotting this function, as we do in Fig.~\ref{IPlots}, for the present case and also for the case of the hyperbolic
tangent velocity profile discussed below.  This figure shows
that these limits are reached very rapidly as a function of $x$, 
approximately justifying our replacement of $\mathfrak{I}(x,y)$ by
its limiting values for $x>0$ and $x<0$.  Defining 
the average energy
$\bar{E}=(E_{1}+E_{2})/2$,
this approximation leads to:
\bea
\label{eq.approx term}
&&\mathfrak{R}(E_{1},E_{2})\simeq -\frac{1}{4\pi^{2}} \int^{0}_{-\infty}dx
{\rm e}^{i(E_1-E_2)x}\int^{\infty}_{-\infty}\frac{dy}{y}\sin(y\bar{E})
\nonumber
\\
 && -\frac{1}{4\pi^{2}}\int^{\infty}_{0}dx{\rm e}^{i(E_1-E_2) x}\int^{\infty}_{-\infty}dy\frac{\sin(y\bar{E})}{2\sinh(y/2)}.
\eea
In this expression we also took into account the fact that $\mathfrak{I}(x,y)$ is odd in $y$.

\begin{figure}[ht!]
\includegraphics[width=\columnwidth]{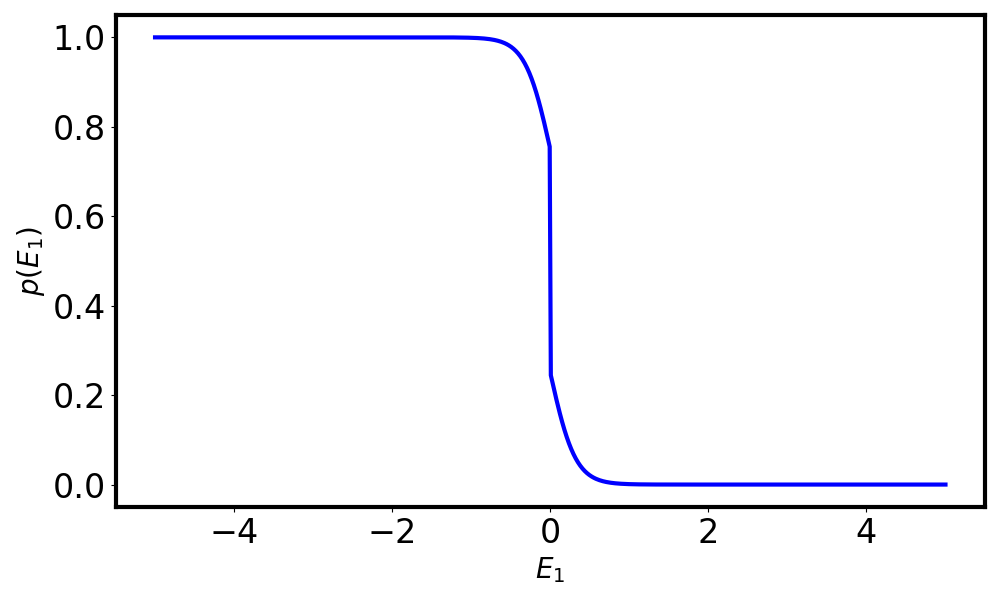}
\caption{
  Plot of the occupation of levels near the Fermi surface, $p(E_1)$, 
  showing the approximate Unruh effect, with a depletion of negative
  energy states and an increase of positive energy states.   Here, $E_1$ is
  the energy normalized to $\kb T_{\rm U}$ with  the Unruh temperature
  given above in Eq.~(\ref{eq:unruhTemp}).
}
\label{Fermisurfaceplot}
\end{figure}

After evaluating the remaining integrals, we arrive at 
\bea
\nonumber 
&&\mathfrak{R}(E_{1},E_{2}) \simeq  -\frac{1}{4}\Big[\sgn(\Ebar) + \tanh(\pi \Ebar)\Big]
\delta(E_1-E_2 ) 
\\
&& \qquad \qquad 
+\frac{i}{4\pi} \Big[\sgn(\Ebar) - \tanh(\pi \Ebar)\Big]
\curP\frac{1}{E_1-E_2 }.
\eea
Upon combining this with Eq.~(\ref{eq.opavg+corr}) we get 
\bea
&&\langle d_{>E_1+}^\dagger d_{>E_2+}\rangle
= \delta(E_1-E_2)p(E_1)\nonumber 
\\
&&
+ \frac{i}{4\pi} \Big[\sgn(\Ebar) - \tanh(\pi \Ebar)\Big]
\curP\frac{1}{E_1-E_2},
\label{eq:finalAverage}
\eea
where the prefactor in the first term is equal to 
\bea
p(E_1) &=& \frac{1}{2}-\frac{1}{4}\sgn(E_1) - \frac{1}{4}
\tanh(\pi E_1) 
\nonumber 
\\
&=&
\begin{cases} \frac{1}{2}\nf(E_1,T_{\rm U}) ,
  & \text{for  $E_1>0$,}\cr
 1-\frac{1}{2}\nf(-E_1,T_{\rm U}) ,
 & \text{for  $E_1<0$,}
 \end{cases} 
 \label{eq:prefactor}
\eea
where in the second line we reintroduced dimensionful quantities that
were previously set to unity, to show that the function $p(E_1)$ (that is
approximately the fermion occupation at energy $E_1$) after the quench
is indeed characterized by the Unruh temperature $T_{\rm U}$ 
defined in  Eq.~(\ref{eq:unruhTemp}).

We see that, while the sigmoid velocity profile does not lead to an 
{\em exact\/} Unruh effect, we do see a modified 
Unruh effect if we can regard the first term of Eq.~(\ref{eq:finalAverage})
as being dominant (this is reasonable since it is proportional to
an energy delta function) and if the approximations leading to this
result are valid.  
The function $p(E_1)$, plotted in Fig.~\ref{Fermisurfaceplot}, shows a 
creation of particles for $E_1>0$ and 
depletion of particles (or creation of holes) for $E_1<0$.  This is
qualitatively consistent with the exact Unruh effect (although reduced in magnitude).  We emphasize that
this is not a true thermal state, since the system is still in the zero temperature Minkoswki vacuum.

\subsection{Hyperbolic Tangent Profile}
In this section we repeat the preceding calculations for a 
velocity profile that is qualitatively similar to the 
sigmoid case, which is the hyperbolic tangent (tanh) velocity function:
\be
v(x) = v\tanh\left(\frac{|x|}{\lambda} \right) .
\ee
The corresponding single-particle Hamiltonian is (taking $v=\lambda = \hbar = 1$
as previously):
\begin{gather}
     H = \sqrt{\tanh|x|}{(-i\partial_{x})}\sqrt{\tanh|x|}.
\end{gather}
Focusing on $x>0$, we find the eigenfunctions at energy $E$ to be:
\begin{gather}
 \psi^{\phdag}_{>E\alpha}(x) = \sqrt{\frac{\coth(x)}{2\pi }}e^{\alpha i E\ln(\sinh(x))}\hat{\chi}^{\phdag}_{\alpha}
\end{gather}
For this profile, we also found the approach of evaluating the momentum integral first to assist in simplifying the analysis.  Indeed,
it turns out the steps are almost identical to the sigmoid case, also
leading to Eq.~(\ref{eq.opavg+corr}) with the function 
$\mathfrak{R}(E_{1}, E_{2})$ still
given by Eq.~(\ref{eq:approxR}) but with a different integrand
$\mathfrak{I}(x, y)$ arising after the variable change 
\begin{gather}
x^{\phdag}_{1}=\sinh^{-1}(e^{x+y/2}), x^{\phdag}_{2}=\sinh^{-1}(e^{x-y/2}).
\end{gather}
The explicit form for $\mathfrak{I}(x, y)$ in the tanh case is: 
\bea
&&\mathfrak{I}(x,y)=\frac{e^{x/2}}{(2[\cosh(2x)+\cosh(y/2)])^{1/4}}\nonumber 
\\
&& \qquad\qquad \times\curP
\frac{1}{\ln\left(\frac{\sqrt{e^{y}+e^{-2x}}+e^{y/2}}{\sqrt{e^{-y}+e^{-2x}}+e^{-y/2}}\right)}.
\label{eq:Itanh}
\eea
Although this function is different from the sigmoid case, the limiting
behavior as a function of $x$ (at fixed $y$) is identical, again
given by Eq.~(\ref{eq:FutureReference}) as illustrated by the solid blue curve of Fig.~\ref{IPlots}.  Thus, a sudden 
quench to the tanh profile gives (within the same approximation) the same result 
Eq.~(\ref{eq:finalAverage}), with the approximate occupation $p(E_1)$ again
given by Eq.~(\ref{eq:prefactor}).  

Having established an approximate Unruh effect within two similar 
models of the post-quench velocity profile, in the next section
we study the real-space spatial extent of the emergent Unruh particles.  

\section{Spatial extent of the Unruh effect}
\label{sec:extent}
We have seen that a quench to an
inhomogeneous velocity profile that is linear at small 
$|x|\ll \lambda$ and constant at large $|x|\gg \lambda$ 
leads to an approximate Unruh effect characterized by 
the creation of positive energy electrons and of negative 
energy holes that looks approximately like a thermal
state.  To gain additional insight into this state, 
in this section we study the spatial extent (or local density) 
of the 
induced positive energy fermions.

We define the local density of positive energy fermions:
\bea
&&\Delta n(x) =\sum_{\alpha,\beta} \int_{0}^{\infty} dE_{1}
\int_{0}^{\infty} dE_{2}\psih^{\dagger}_{>E_{1}\alpha }(x)\psih^{\phdag}_{>E_{2}\beta}(x)\nonumber 
\\
&&\qquad \qquad\qquad \qquad  \times 
\langle d^{\dagger}_{>E_{1}\alpha} d_{>E_{2}\beta}^{\phdag}\rangle.
\eea
Plugging in our result for the final occupation 
Eq.~(\ref{eq:finalAverage}) that applies to both cases,
we find the contribution due to the second line vanishes,
leaving
\be
\Delta n(x) = \sum_{\alpha = \pm}\int_{0}^{\infty} dE
\psih^{\dagger}_{>E\alpha}(x)\psih^{\phdag}_{>E\alpha }(x)\frac{1}{2}\nf(E,T_U).
\ee
Interestingly, the product of wavefunctions in the integrand
of this expression is, generally, simply proportional to 
$1/v(x)$.  To see this, we note that eigenfunctions for 
Eq.~(\ref{eq:flatHam}) take the form (with $\hbar = 1$):
\be
\psih_{>E\alpha}(x) = \hat{\chi}_\alpha \frac{c}{\sqrt{v(x)}}
{\rm e}^{i\alpha E \int_{x_0}^x dx'\frac{1}{v(x')}},
\ee
with  $x_0$ an arbitrary initial position and $c$ a normalization factor (equal to $1/\sqrt{2\pi}$ in both cases), immediately implying 
\be
\label{Eq:localdensitychange}
\Delta n(x) =
\frac{|c|^2}{v(x)}
\int_{0}^{\infty} dE \nf(E,T_U) = \frac{|c|^2}{v(x)} \kb T_U \ln 2.
\ee
This result (plotted as dashed curves in Fig.~\ref{fig:LFD}, with dimensionful quantities set to unity)
shows that the positive-energy fermions 
(i.e., the $E_1>0$ regime of Fig.~\ref{Fermisurfaceplot})
emerging from the approximate Unruh effect induced by
a spatially-varying velocity $v(x)$ are localized, spatially,
near $x=0$ where $v(x)\to 0$. The $E_1<0$ 
regime of Fig.~\ref{Fermisurfaceplot}) shows a similar
depletion of fermions for negative energies; these will 
give a local density change that is exactly the opposite of 
Eq.~(\ref{Eq:localdensitychange}).  Although the total
local charge density is therefore unchanged, the energy
dependence can likely be probed by energy-sensitive
probes like scanning tunneling microscopy.  We leave the investigation of
further experimental probes of this state for future work.

\section{Concluding Remarks}
\label{sec:concl}
We have studied the Unruh effect in 
a model of one-dimensional Dirac fermions with a 
controllable local Dirac velocity $v(x)$ that (by assumption) can be spatially
uniform or spatially nonuniform.  The uniform case $v$ case corresponds
to a conventional theory of 1D Dirac fermions, with a \lq\lq Minkowski\rq\rq\
vacuum ground state.  

The  Unruh effect occurs after a sudden quench to an inhomogeneous
velocity profile, with  the system occupation characterized
by an  Fermi distribution  with an emergent temperature
$T_{\rm U}$.  After first studying the ideal case with
$v(x)\propto |x|$ (in which the final system is connected to
Dirac fermions in Rindler spacetime), we turned to the case
of more experimentally-realizable velocity profiles that are linear at small $|x|$ and homogeneous for large $|x|$ (beyond a length scale $\lambda$).  
The latter cases led to an {\em approximate} Unruh effect, with a distribution of fermions also
characterized by a Fermi distribution.  As a signature of this approximate
Unruh effect, we studied the spatial profile, as a function of position,
of the number of positive-energy excitations, finding it to be localized
near $x=0$ with a profile proportional to the reciprocal of the final 
velocity profile.

Our work shows that the physics of the Unruh effect can emerge in
condensed matter systems that only approximately realize the conditions
of an ideal Unruh effect.  Future possible directions for
study include computing other observables (including transport
and thermodynamic properties as well as probes
like tunneling and photoemission), generalizing to the case of
a finite temperature initial state, and studying 
the predicted statistics inversion of the Unruh 
effect~\cite{Takagi,Ooguri} that is connected to Huygens' principle
for wave motion.

\section{Acknowledgments}
 DES acknowledges support from the National Science Foundation under Grant PHY-2208036.    DES performed part of this work  at the Aspen Center for Physics, which is supported by National Science Foundation grant PHY-2210452.

\end{document}